\documentclass[12pt,preprint]{aastex}

\newcommand{\ld}{\lambda/D}
\newcommand{\rhonaught}{\rho_{\mbox{\scriptsize iwd}}}
\newcommand{\rhoone}{\rho_{\mbox{\scriptsize owd}}}
\newcommand{\hoverra}{\alpha(r)}    
\newcommand{\hoverrb}{\alpha(r)}  
\newcommand{\hovertwor}{\frac{\alpha(r)}{2}}  

\usepackage[intlimits]{amsmath}

\newcommand{\pz}{\phantom{0}}  

\shorttitle{Starshape Masks}
\shortauthors{Vanderbei et al.}

\begin{document}


\title{Circularly Symmetric Apodization via Starshaped Masks}


\author{Robert J. Vanderbei}
\affil{Operations Research and Financial Engineering, 
Princeton University}
\email{rvdb@princeton.edu}

\author{David N. Spergel}
\affil{Astrophysical Sciences, Princeton University}
\email{dns@astro.princeton.edu}

\and

\author{N. Jeremy Kasdin}
\affil{Mechanical and Aerospace Engineering, Princeton University}
\email{jkasdin@princeton.edu}


\begin{abstract}
In \cite{VSK02}, we introduced a class of shaped pupil masks,
called spiderweb masks, that produce point spread functions having annular
dark zones.  With such masks, a single image can be used to probe
a star for extrasolar planets.  In this paper, we introduce a new
class of shaped pupil masks that also provide annular dark zones.  We
call these masks starshaped masks.
Given any circularly symmetric apodization function, we show how to
construct a corresponding starshaped mask that has the same point-spread
function (out to any given outer working distance) as obtained
by the apodization.  

\end{abstract}



\section{Introduction}
\label{sec:intro}

With more than 100 extrasolar Jupiter-sized
planets discovered in just the last decade, 
there is now great interest in discovering
and characterizing Earthlike planets.  To this end, NASA is planning to
launch a space-based telescope, 
called the {\em Terrestrial Planet Finder (TPF)},
sometime in the middle of the next decade.  This telescope, which
will ultimately be either an interferometer or a coronagraph, will be
specifically designed for high-contrast imaging.  Earlier studies 
(\citet{ref:Brown}) 
indicate that a $4$m class coronagraph ought to be able to discover
about 50 extrasolar Earth-like planets if the telescope can provide contrast
of $10^{-10}$ at a separation of $3\ld$ and that a $4 \times 10$m class
telescope ought to be able to discover about 150 such
planets if it can provide the same contrast at a separation of $4\ld$.

One of the most promising design concepts for high-contrast imaging
is to use pupil masks for diffraction
control.
Early work in this direction 
(see \citet{
ref:Brown,ref:Spergel,KVSL02b,KVSL02}) 
has focused on optimizing masks that are not rotationally symmetric
and thus provide the desired contrast only in a narrow annular sector around a
star, but at fairly high throughput.  
A full investigation of a given star then requires multiple images
where the mask is rotated between the images so as to image
all around the star.  In a recent paper (\cite{VSK02}), 
we proposed a new class of rotationally
symmetric pupil masks, which are attractive because they do not require
rotation to image around a star.  Such masks consist of concentric rings 
supported by many ultra-thin pie-shaped spiders.  
We call these masks {\em spiderweb} masks.  We showed that
they can provide the desired contrast with a reasonable amount of
light throughput.  

In this paper, we consider a similar family of binary masks.  These
masks consist only of spiders---there are no concentric rings.  Instead,
the desired contrast is obtained by carefully controlling the width of the
spiders as a function of radius.  We call these masks {\em starshaped} masks
since the resulting designs look like many-pointed stars.

The paper is organized as follows.  In the next section, 
we briefly review the relationship between pupil-plane apodization 
and the corresponding image-plane electric field and point-spread function.
Then in Section \ref{sec:stars}, we show how to make starshape binary masks
for which the first term in a Bessel expansion of the electric field
matches the corresponding electric field associated with 
any given apodization.  We show that for a sufficiently large
number of star-points, the remaining terms have negligible contribution to
the electric field in the annulus of interest.
Finally, in Section \ref{sec:apod} we show how to compute various ``optimal''
apodizations.  In particular, we show that the best of these optimal
apodizations is zero-one valued and in fact corresponds to the optimal
spiderweb masks presented in \citet{VSK02}.

\section{Apodization}
\label{sec:masks}

The image-plane electric field produced by 
an on-axis point source and an apodized aperture defined
by a circularly-symmetric apodization function $A(\sqrt{x^2 + y^2})$ is given
by 
\begin{equation}
    E(\xi,\zeta) = 
    \iint_S e^{-2 \pi i (x \xi + y \zeta )} A\left(\sqrt{x^2+y^2}\right) dx dy, 
\end{equation}
where
\begin{equation}
    S = \displaystyle \left\{ (x,y): \;
    		          0 \le r(x,y) \le 1/2, \;
                          \theta(x,y) \in \left[0, 2\pi \right]
		       \right\},
\end{equation}
and $r(x,y)$ and $\theta(x,y)$ denote the polar coordinates associated with
point $(x,y)$.  
Here, and throughout the paper, $x$ and $y$ denote coordinates in the pupil
plane measured in units of the aperature $D$ and $\xi$ and $\zeta$ denote
angular (radian) deviations from on-axis measured in units of wavelength
over aperture ($\lambda/D$) or, equivalently, physical distance in the image
plane measured in units of focal-length times wavelength over aperture 
($f \lambda/D$).

For circularly-symmetric apodizations,
it is convenient to work in polar
coordinates.  To this end, let $r$ and $\theta$ denote polar coordinates in
the pupil plane and let $\rho$ and $\phi$ denote the image plane coordinates:
\begin{equation}
    \begin{array}{rclrcl}
        x & = & r \cos \theta \qquad \qquad   \xi & = & \rho \cos \phi \\
        y & = & r \sin \theta \qquad \qquad \zeta & = & \rho \sin \phi .
    \end{array}
\end{equation}
Hence,
\begin{eqnarray}
 x \xi + y \zeta &=& r \rho (\cos \theta \cos \phi + \sin \theta \sin \phi)\\
                 &=& r \rho \cos(\theta-\phi)  .
\end{eqnarray}
The electric field in polar coordinates depends only on $\rho$ and is given by
\begin{eqnarray}
    E(\rho) 
    &=&  \label{eq1}
    \int_0^{1/2} \int_0^{2\pi} 
        e^{-2 \pi i r \rho \cos(\theta-\phi)} A(r) 
    r d\theta dr, \\
    &=&  \label{eq2}
    2 \pi \int_0^{1/2} J_0(2 \pi r \rho) A(r) r dr,
\end{eqnarray}
where $J_0$ denotes the $0$-th order Bessel function of the first kind.
Note that the mapping from apodization function $A$ to electric field $E$
is linear.  Furthermore, the electric field in the image plane is real-valued
(because of symmetry).  

The {\em point spread function} (psf) is the square of the electric field in
the image plane.
The contrast requirement is that the psf in the dark region be $10^{-10}$ of
what it is at its center.  Because the electric field is real-valued, it
is convenient to express the contrast requirement in terms of it rather than
the psf, resulting in a field requirement of $\pm 10^{-5}$.

A second requirement is high throughput.  The most natural measure of
throughput is the area under the main lobe of the psf:
\begin{equation} \label{300}
    {\mathcal T}_{\rm{Airy}} = \int_0^{\rhonaught} E^2(\rho) 2\pi\rho d\rho ,
\end{equation}
where $\rhonaught$ denotes the location of the first null of the psf.
We call this measure the {\em Airy-throughput}.
Two other measures of throughput are relevant to our discussion.
The {\em total throughput} of an apodization is the integral of the psf over
the entire image plane.  By Parseval's theorem, the total throughput is also
the integral of the square of the apodization:
\begin{equation}
    {\mathcal T}_{\rm{total}} =
    \int_0^{\infty} E^2(\rho) 2\pi\rho d\rho
    =
    \int_0^{1/2} A^2(r) 2\pi r dr .
\end{equation}
The electric field at $\rho = 0$,
\begin{equation}
    E(0) = \int_0^{1/2} A(r) 2 \pi r dr ,
\end{equation}
provides another measure of the ``central'' throughput of the apodization, 
since its square is the peak throughput density at the center of the Airy disk.
If $A()$ is zero-one valued, then the apodization can be realized as a mask.
In this case, $E(0)$ is precisely the open area of the mask (and
is also the total-throughput).  For this reason, we call $E(0)$ 
the {\em pseudo-area} of the apodization even when the
apodization is not zero-one valued.

\section{Starshape Masks}
\label{sec:stars}

\begin{figure}
\begin{center} 
\includegraphics[width=2.0in]{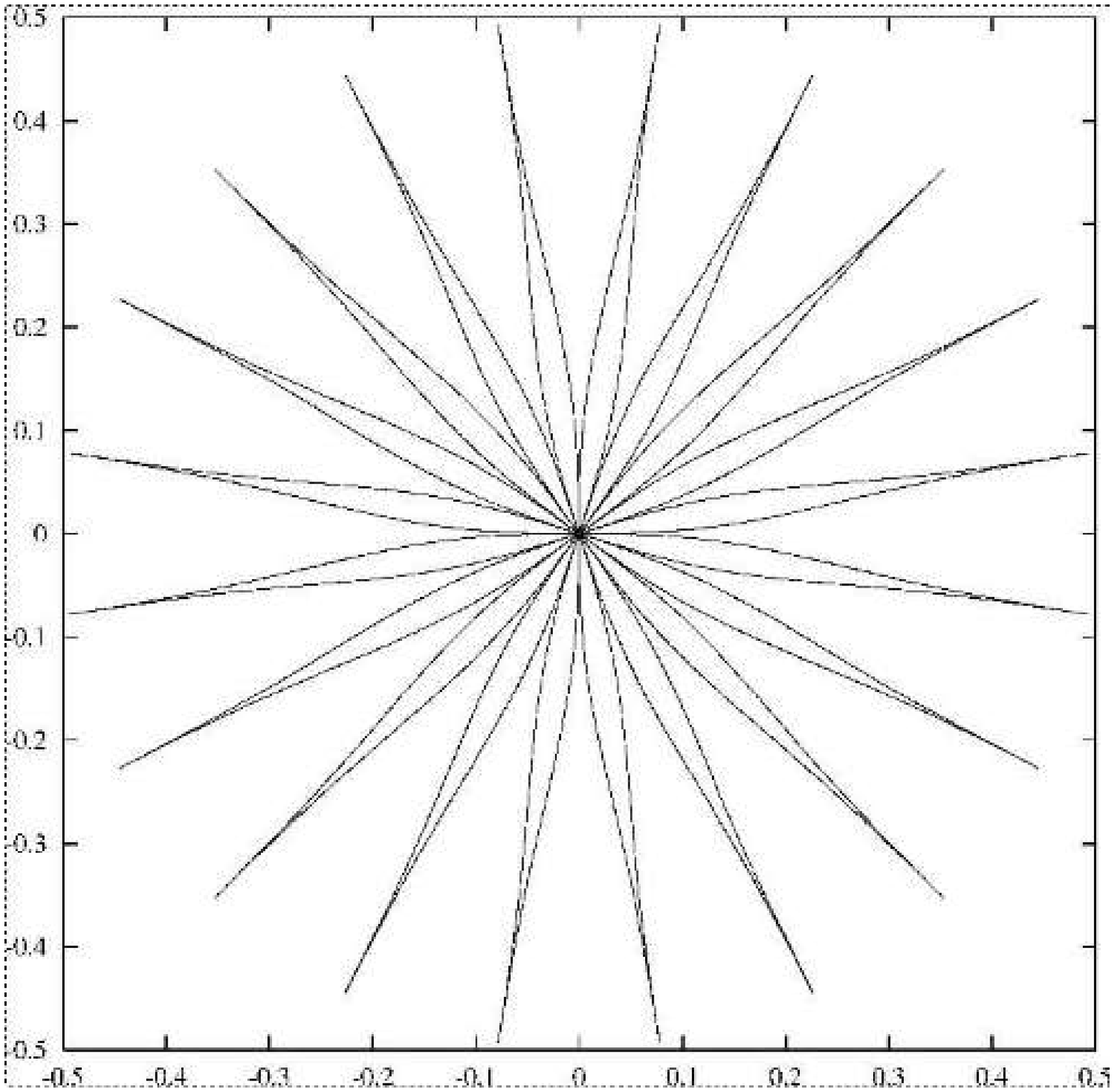}
\end{center}
\caption{A $20$-point starshape mask corresponding to the
apodization shown in Figure \ref{fig:fig7}. }
\label{fig:fig1}
\end{figure}

In this section, we study binary mask approximations to any apodized pupil.
The masks we consider are {\em starshaped}---see Figure \ref{fig:fig1}.
The opening in an $N$-point starshaped mask can be described 
mathematically by a set $S$ given by:
\begin{equation}
    S = \displaystyle \left\{ (x,y): \;
    		          0 \le r(x,y) \le 1/2, \;
                          \theta(x,y) \in \Theta 
		       \right\},
\end{equation}
\begin{equation} \label{102}
    \Theta = \bigcup_{n=0}^{N-1} 
		   \left[
			   \frac{2 \pi n}{N}     + \hovertwor,
			   \frac{2 \pi (n+1)}{N} - \hovertwor
		   \right] ,
\end{equation}
where
$\hoverra$ denotes the width in radians of a ``vane'' and the notation
$[a,b]$ denotes the interval on the real line from $a$ to $b$.
Note that the shape
of each point of the star is determined by the function $\alpha(r)$.
Our aim is to determine choices for this function that will yield
an image-plane psf matching the psf corresponding to any given apodization.

The electric field, expressed in polar coordinates, associated with
starshape mask $S$ is given by
\begin{equation} \label{eq3}
    E(\rho,\phi) 
    =
    \displaystyle
    \iint_S e^{-2 \pi i r\rho \cos(\theta-\phi)} r dr d\theta .
\end{equation}

The integral in equation \eqref{eq3} can be expressed in terms
of Bessel functions using the Jacobi-Anger expansion (see, e.g., 
\citet{AW00} p. 681):
\begin{equation}
    e^{i x \cos \theta} 
    =
    \sum_{m=-\infty}^{\infty} i^m J_m(x) e^{i m \theta} .
\end{equation}
Substituting into \eqref{eq3}, we get:
\begin{eqnarray}
    E(\rho,\phi)
    &=&
    \displaystyle
    \iint_S 
       \sum_m i^m J_m(-2 \pi r \rho) e^{im(\theta-\phi)}  
    r dr d\theta \\
    &=&
    \displaystyle
    \int_0^{1/2} 
       \sum_m i^m J_m(-2 \pi r \rho) e^{-im\phi} 
       \left( \int_{\Theta} e^{i m \theta} d\theta \right)
    r dr 
\end{eqnarray}
The integral over $\Theta$ is easy to compute:
\begin{eqnarray}
    \int_{\Theta} e^{i m \theta} d \theta
    &=&
    \sum_{n=0}^{N-1} 
    \int_{\frac{2 \pi n}{N}+\hovertwor}^{\frac{2 \pi (n+1)}{N}-\hovertwor} 
        e^{im\theta} d\theta
    \\
    &=&
    \left\{
	\begin{array}{ll}
            2 \pi - N\hoverrb & \quad m = 0 \\
	                       - \frac{2}{j} \sin(jN\hovertwor)
			       	 & \quad m = jN, \; j \ne 0 \\
			       0 & \quad \mbox{otherwise} .
	\end{array}
    \right.
\end{eqnarray}
Substituting this result into \eqref{eq3}, yields
\begin{eqnarray}
    E(\rho,\phi)
    &=&
    \int_0^{1/2} J_0(-2\pi r \rho)(2\pi - N\hoverrb) r dr \\
    && - \sum_{j \ne 0}
    \int_0^{1/2} i^{jN} J_{jN}(-2\pi r \rho)e^{-ijN\phi} 
	                            \frac{2}{j} \sin\left(jN\hovertwor\right)
				    r dr .
\end{eqnarray}
Lastly, suppose that $N$ is even 
and use the fact that $J_{-m}(x) = J_m(-x) = (-1)^m J_m(x)$ to get the
following expansion for the electric field:
\begin{eqnarray}
    E(\rho,\phi)
    &=&
    2 \pi \int_0^{1/2} J_0(2\pi r \rho) (1 - \frac{N}{2\pi}\hoverrb) r dr \\
    && - 4 \sum_{j=1}^{\infty} 
        \int_0^{1/2} J_{jN}(2\pi r \rho)\cos(jN(\phi-\pi/2)) 
	                       \frac{1}{j} \sin(jN\hovertwor)
				rdr .
\end{eqnarray}

The first term, involving the integral of $J_0$,
is identical to the formula for the electric
field for an apodized aperture with  
\begin{equation} \label{101}
    A(r) = 1 - \frac{N}{2\pi} \hoverrb .
\end{equation}
Hence, by putting
\begin{equation}
    \alpha(r) = \frac{2\pi}{N} \left(1 - A(r)\right),
\end{equation}
we can make the first term match any circularly symmetric apodization.

Furthermore, 
for large $N$, the effect of the higher-order Bessel terms becomes negligible
for small $\rho$.  Indeed, for $z \le \sqrt{4(m+1)}$, 
\begin{equation}
    0 \le J_m(z) \le \frac{(z/2)^{m+1}}{(m+1)!} ,
\end{equation}
(which itself follows easily from the alternating Taylor series expansion
of the $m$-th Bessel function: 
$J_m(z) = \sum_{l=0}^{\infty} (-1)^l (z/2)^{2l+m}/(l!(m+l)!)$).
From this we use the Schwarz inequality to estimate the magnitude of the
effect of the higher-order terms:
\begin{equation}
\setlength{\arraycolsep}{0in}
\begin{array}{rcl}
    \displaystyle
    \left| \phantom{\int_0^{1/2}}4 \sum_{j=1}^{\infty} 
        \int_0^{1/2} J_{jN}(2\pi r \rho)\cos(jN(\phi \right.&-&\left. \pi/2)) 
			       \displaystyle
	                       \frac{1}{j} \sin(jN\hovertwor)
				rdr
				\phantom{\int_0^{1/2}}
				\right| \\
	&\le& \; \displaystyle
	      \frac{1}{2} \sum_{j=1}^{\infty} J_{jN}(\pi \rhoone) \\
	&\le& \; \displaystyle
	      \frac{1}{2} \sum_{j=1}^{\infty} 
	      \frac{1}{(jN+1)!} \left( \frac{\pi \rhoone}{2} \right)^{jN+1} ,
\end{array}
\end{equation}
for $\rhoone \le \sqrt{4(N+1)}/\pi$.
Since the last bound is dominated by $e^{\pi \rhoone/2}/2$, it follows from
the dominated convergence theorem that this last bound tends to zero as $N$
tends to infinity.

\begin{figure}
\begin{center} 
\includegraphics[width=3.0in]{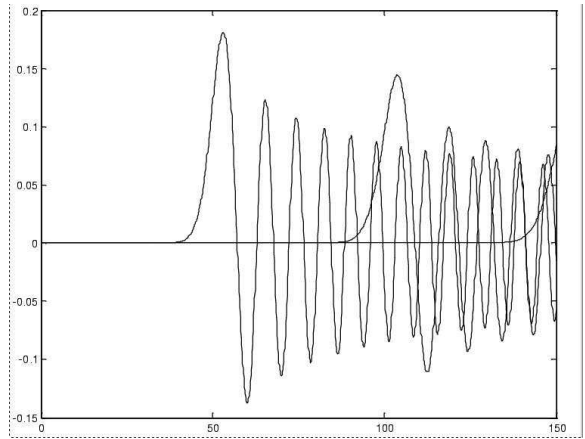}
\end{center}
\caption{The Bessel functions $J_{50}$, $J_{100}$, and $J_{150}$.
They first reach $10^{-5}$ at $35.2$, $81.0$, and $128.1$, respectively.
}
\label{fig:fig8}
\end{figure}

The convergence to zero of the terms in the sum on $j$ is very fast.
In fact, if $N$ is set large enough such that 
$\max_{0 \le z \le \pi \rhoone} J_N(z) \le 10^{-5}$, 
then the $j=1$ term dominates the sum of all the higher-order terms and
is itself dominated by the $J_0$ term.  
Figure \ref{fig:fig8} shows a plot of
$J_{50}$, $J_{100}$, and $J_{150}$.  These three Bessel functions first reach
$10^{-5}$ at $z=35.2$, $81.0$, and $128.1$, respectively.
This suggests that a contrast level of $10^{-10}$ can be preserved out to
$35 \lambda/D$ using about $N=50$ vanes, out to $80 \lambda/D$ using about
$N=100$ vanes, and out to $120 \lambda/D$ using about $N=150$ vanes.
These estimates ignore the contribution from higher-order terms.  In practice,
more vanes are required to preserve the desired contrast level to these outer
working distances.


Within the discovery zone, the electric fields 
resulting from the apodization and the starshaped
mask agree.   Hence, the Airy-thoughputs and pseudo-areas also agree.
But, the total throughput for the mask is larger, often by a factor
of about two.  Indeed, the total throughput of the apodization is
\begin{equation}
    {\mathcal T}_{\text{total,apod}} = \int_0^{1/2} A(r)^2 2 \pi r dr
\end{equation}
whereas, from \eqref{102} and \eqref{101}, it follows that
the total throughput of the mask is
\begin{equation}
    {\mathcal T}_{\text{total,mask}} = \iint_S r dr d\theta 
                              = \int_0^{1/2} A(r) 2 \pi r dr
			      \ge {\mathcal T}_{\text{apod}} 
\end{equation}
(the inequality holds because $0 \le A(r) \le 1$).
So, the mask lets about twice as much light through but the extra
light is necessarily concentrated outside the outer working distance.

\section{Apodizations Optimized for TPF} \label{sec:apod}
There has been much interest in using apodized pupils to achieve
high-contrast imaging, especially recently in the context of TPF studies:
see e.g. \citet{ref:Jacquinot,ref:Indebetouw,ref:Watson,ref:Nisenson}.

The simplest optimization problem involves maximizing the pseudo-area 
subject to contrast constraints.  It
can be formulated as an infinite dimensional linear optimization
problem:
\begin{equation} \label{100}
    \begin{array}{ll}
        \mbox{maximize } & \int_0^{1/2} A(r) 2\pi r dr \\
	\mbox{subject to } &
	    \setlength{\arraycolsep}{0.1em}
	    \begin{array}[t]{rcll}
	    \displaystyle
	        -10^{-5} E(0) \le & E(\rho) & \le 10^{-5} E(0), &
	            \qquad \rhonaught \le \rho \le \rhoone , \\
		  0 \le & A(r)   & \le  1, & \qquad 0 \le r \le 1/2, \\
	    \end{array}
    \end{array}
\end{equation}
where $\rhonaught$ denotes a fixed {\em inner working distance} and $\rhoone$ a
fixed {\em outer working distance}.
%
Discretizing the sets of $r$'s and $\rho$'s and replacing the integrals with
their Riemann sums, 
problem \eqref{100} is approximated by a finite dimensional linear
programming problem,  
which can be solved to a high level of precision (see, e.g., \citet{Van01}).
The numerical solution to this problem reveals that the optimal solution is
zero-one valued.   That is, the optimal apodization is, in fact, a mask
consisting of concentric rings.

A second optimization problem involves maximizing total-throughput:
\begin{equation} \label{130}
    \begin{array}{ll}
        \mbox{maximize } & \int_0^{1/2} A(r)^2 2\pi r dr \\
	\mbox{subject to } &
	    \setlength{\arraycolsep}{0.1em}
	    \begin{array}[t]{rcll}
	    \displaystyle
	        -10^{-5} E(0) \le & E(\rho) & \le 10^{-5} E(0), &
	            \qquad \rhonaught \le \rho \le \rhoone , \\
		  0 \le & A(r)   & \le  1, & \qquad 0 \le r \le 1/2. \\
	    \end{array}
    \end{array}
\end{equation}
An empirical observation is that this more difficult optimization problem
also proves to be tractible and, in all cases tried, provides the same
optimal apodization as was obtained by maximizing pseudo-area.

\begin{figure}
\begin{center} \includegraphics[width=2.5in]{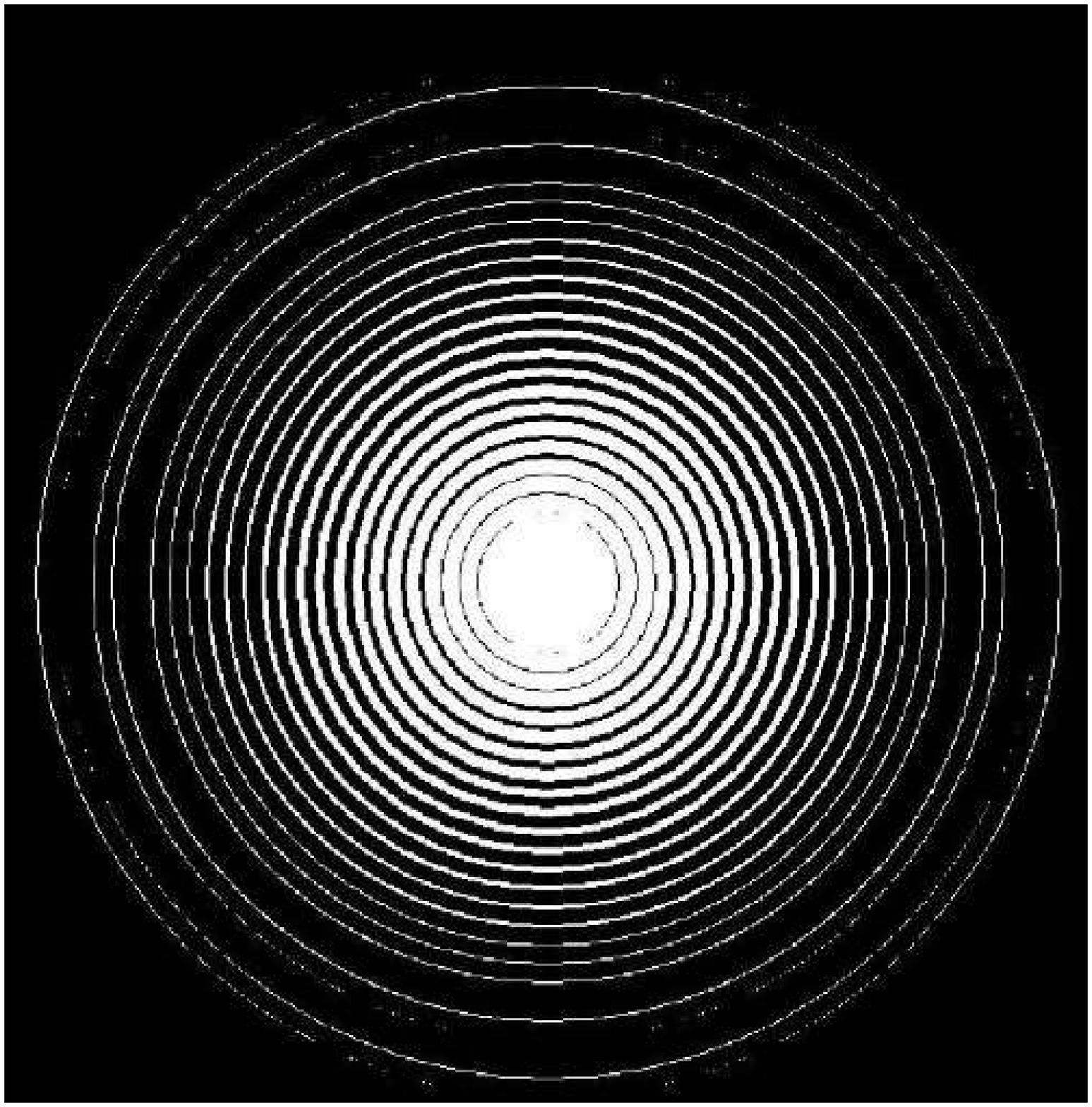} \end{center}
\begin{center} 
\hfill \includegraphics[width=2.0in]{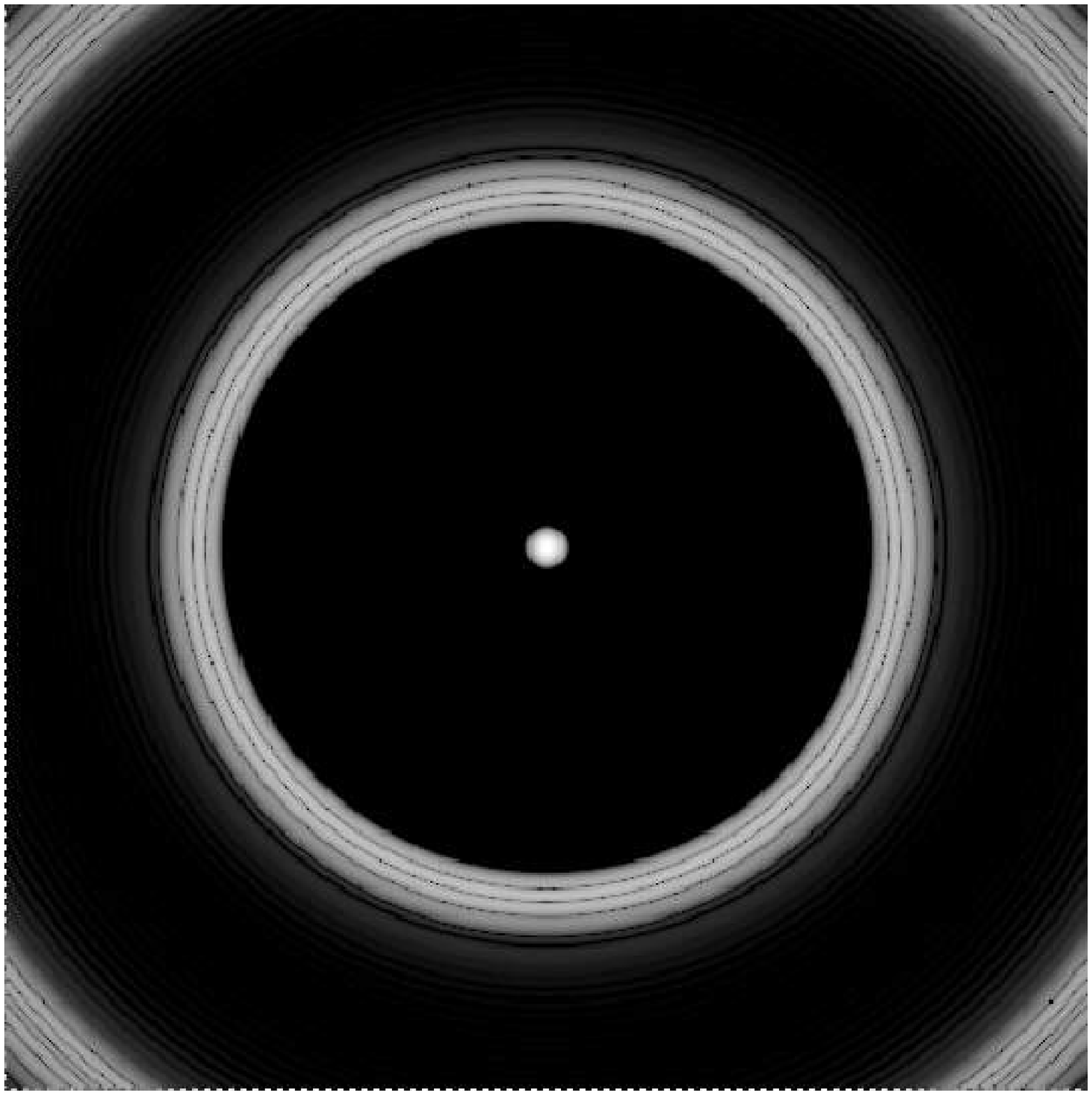}
\hfill \includegraphics[width=2.5in]{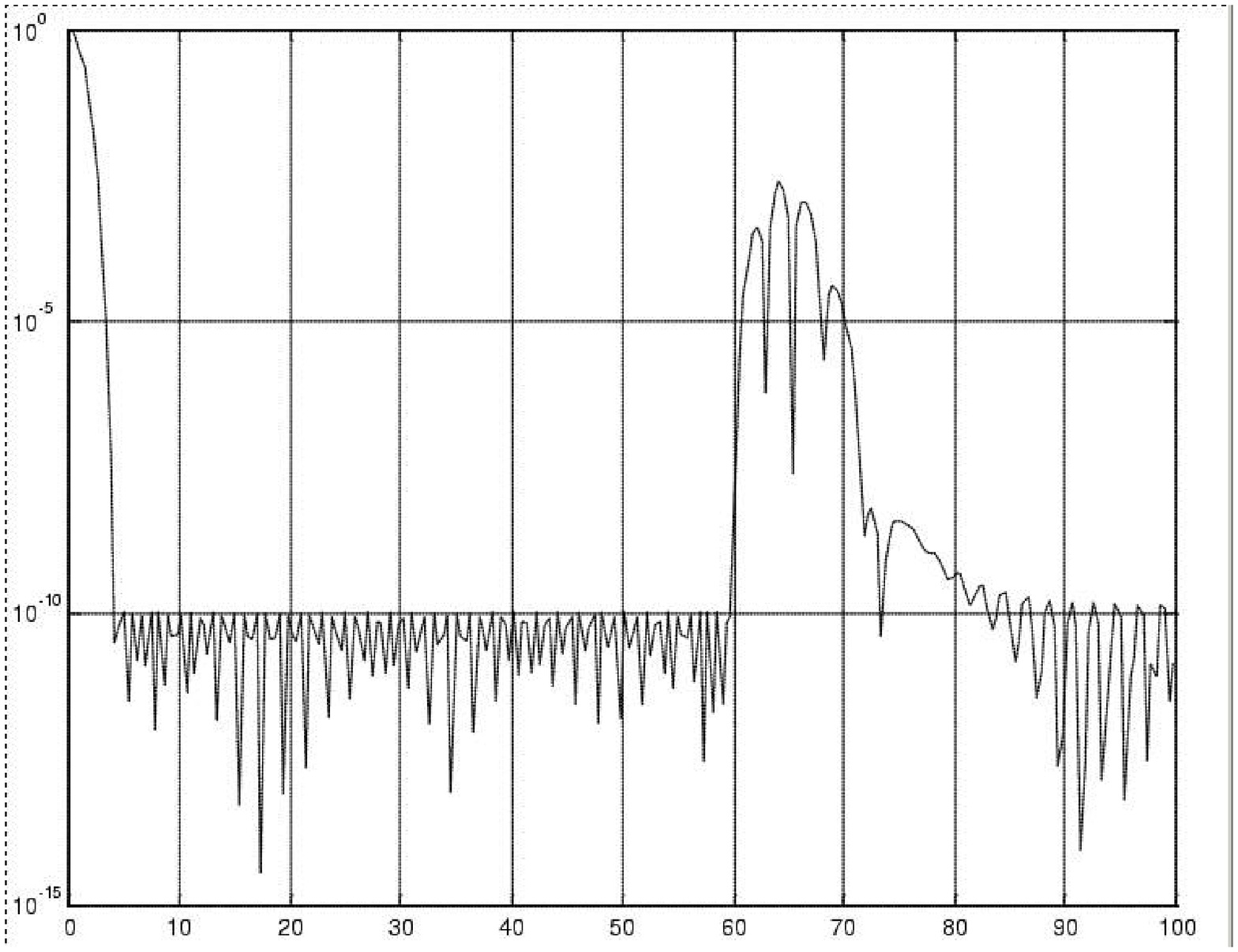}
\hfill ~ 
\end{center}
\caption{{\em Top.} A concentric-ring mask designed to provide high-contrast,
$10^{-10}$, from $\ld = 4$ to $\ld=60$.  Total throughput and pseudo-area are
$17.9\%$.  Airy throughput is $9.37\%$.
{\em Bottom.} The associated psf.
}
\label{fig:fig10}
\end{figure}

The solution obtained for $\rhonaught = 4$ and $\rhoone = 60$ is shown in 
Figure \ref{fig:fig10}.   
The throughput is $17.9\%$ and, since it is a mask, its pseudo-area is the same.
Its Airy-thoughput is $9.37\%$.  (Note: all results are reported as a
percentage of the largest possible value which would result from a completely
open aperture.  That is, all throughputs are divided by the area $\pi (1/2)^2$
and labeled as percentages.)

The starshaped mask associated with this ``apodization'' is precisely
the same concentric-ring mask independent of the number of star points chosen, 
since the star
points disappear entirely.  Of course, this mask cannot be manufactured as
the concentric rings need some means of support.  If one imposes an upper
bound on $A$ in \eqref{100} that is strictly less than unity, then one obtains
a supportable mask with slightly reduced throughput (the amount depending
on the upper bound chosen).  Such a mask is exactly the spiderweb mask
studied in \cite{VSK02}.

In this paper, we are interested in smooth apodizations.
To this end, we add seemingly artificial
smoothness constraints to the optimization problem.
Motivated by the fact that optimal apodizations look qualitatively like
a Gaussian function, we impose smoothness constraints that correspond to the
following conditions:
\begin{eqnarray}
    \log(A)' & \le & 0 \\
    \log(A)'' & \le & 0 .
\end{eqnarray}
The resulting max-pseudo-area optimization problem is:
\begin{equation} \label{120}
    \begin{array}{ll}
        \mbox{maximize } & \int_0^{1/2} A(r) 2\pi r dr \\
	\mbox{subject to } &
	    \setlength{\arraycolsep}{0.1em}
	    \begin{array}[t]{rcll}
	    \displaystyle
	        -10^{-5} E(0) \le & E(\rho) & \le 10^{-5} E(0), &
	            \qquad \rhonaught \le \rho \le \rhoone , \\
		  0 \le & A(r)   & \le  1, & \qquad 0 \le r \le 1/2, \\
		        & A'(r)  & \le  0, & \qquad 0 \le r \le 1/2, \\
		        & A(r) A''(r)& \le A'(r)^2, & \qquad 0 \le r \le 1/2. \\
	    \end{array}
    \end{array}
\end{equation}

\begin{figure}
\begin{center} 
\includegraphics[angle=-90,width=3.0in]{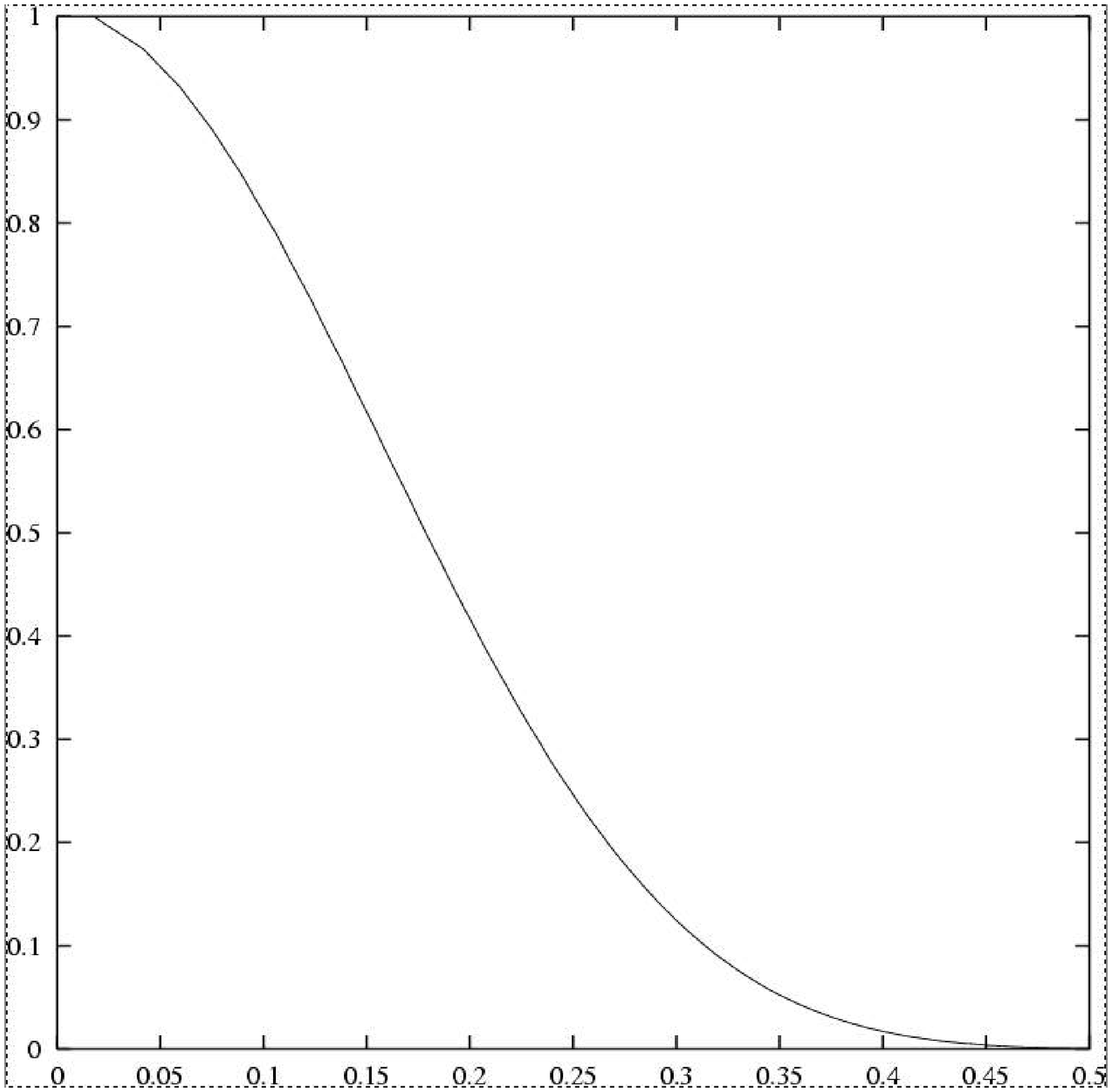}
\includegraphics[angle=-90,width=3.0in]{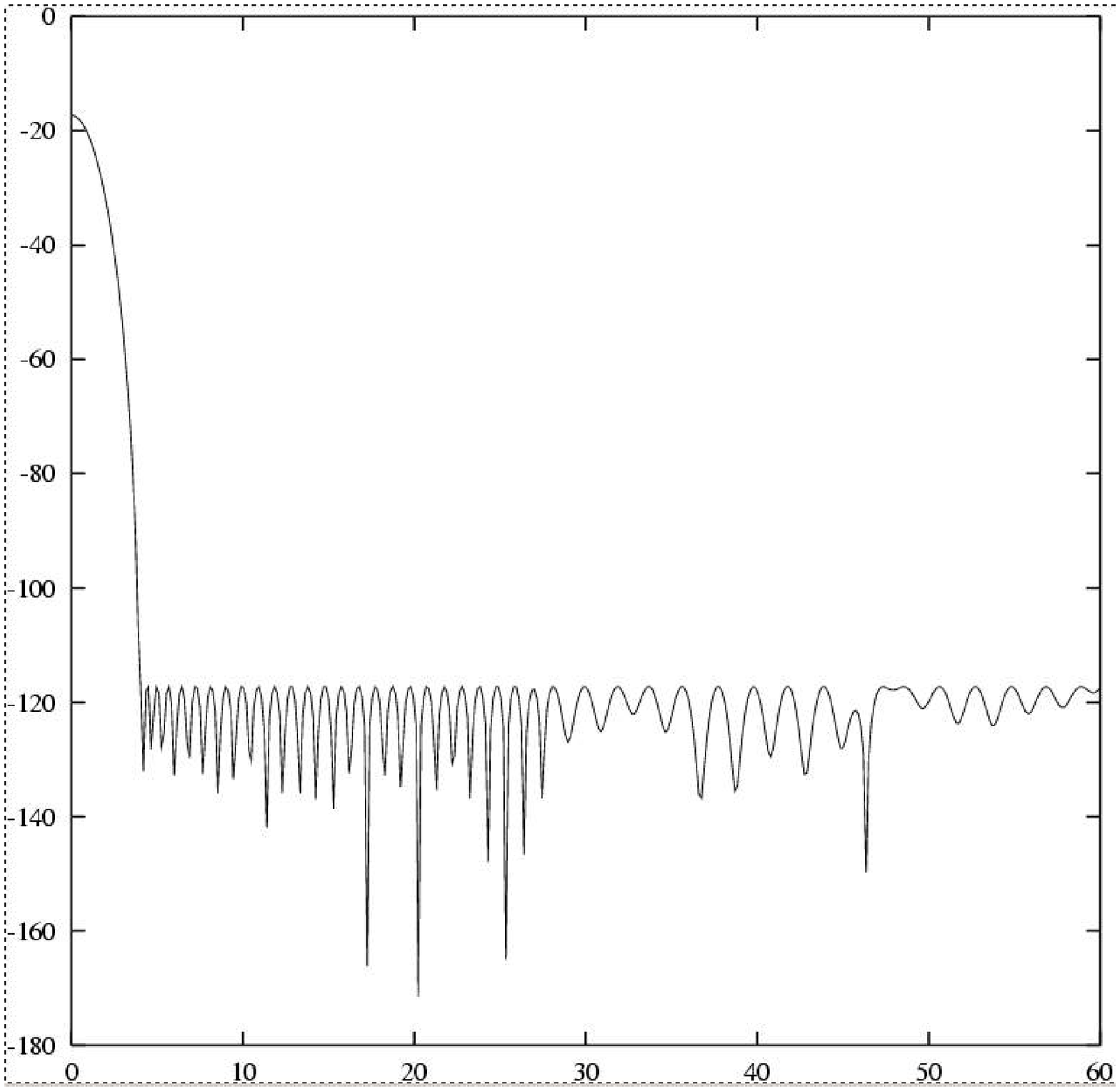}
\end{center}
\caption{The optimal apodization for $\rhonaught = 4$ and $\rhoone = 60$
and the associated psf.  Total throughput is $9.12\%$.  Pseudo-area is
$17.39\%$.  Airy-throughput is $9.09\%$.
}
\label{fig:fig7}
\end{figure}

\begin{figure}
\begin{center} 
\hfill
\includegraphics[width=2.0in]{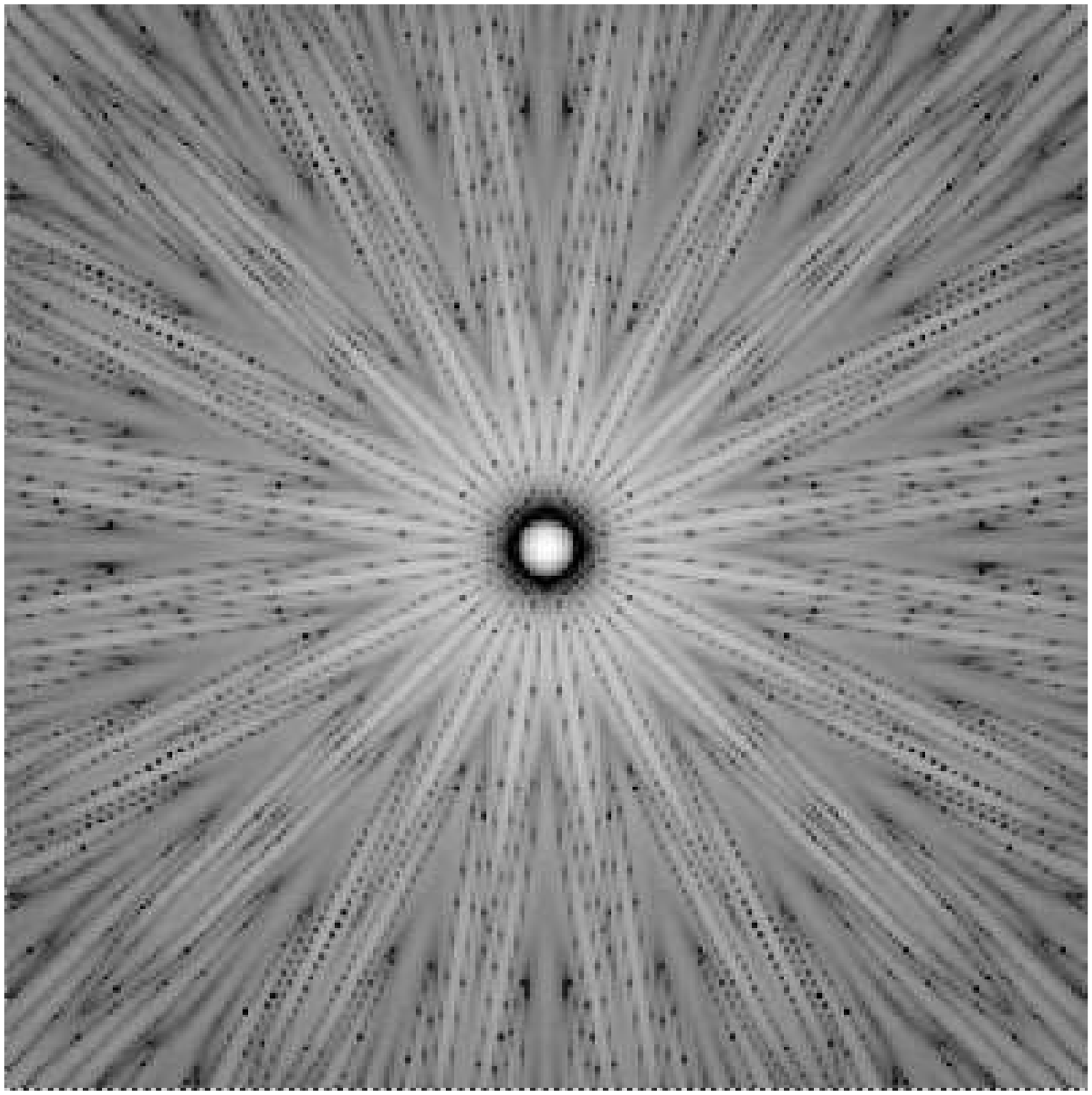}
\hfill
\includegraphics[width=2.5in]{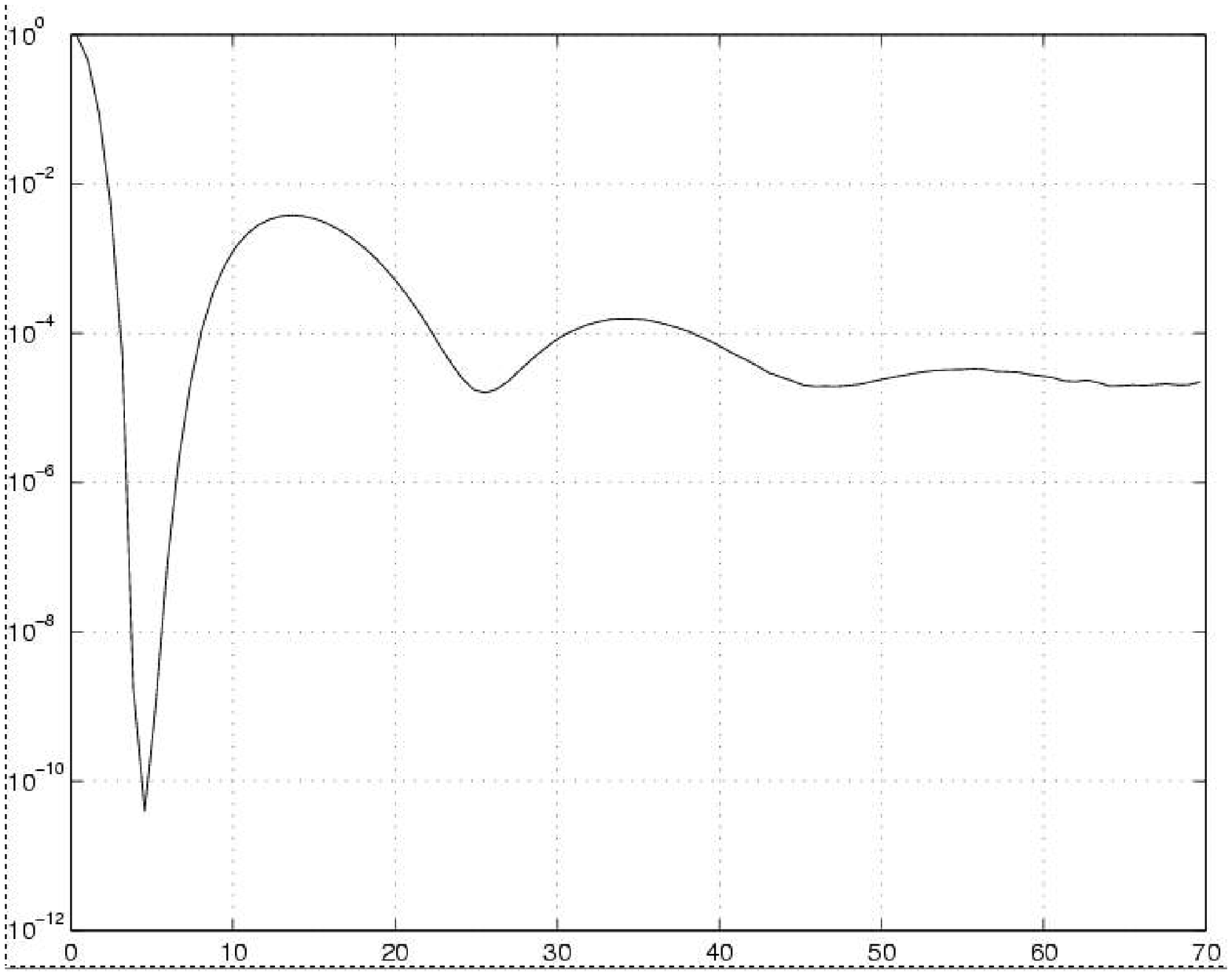}
\hfill ~ \\
\hfill
\includegraphics[width=2.0in]{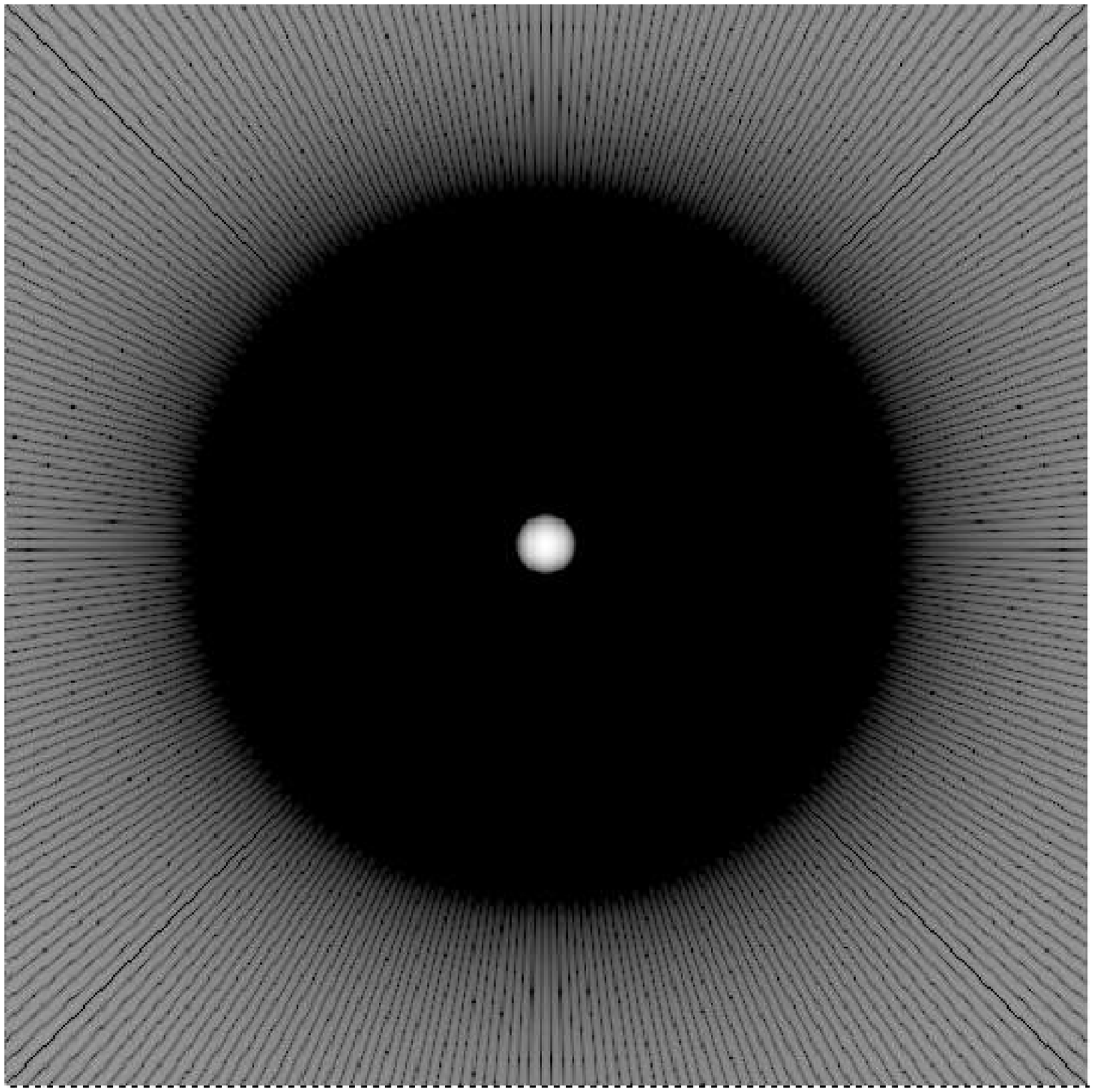}
\hfill
\includegraphics[width=2.5in]{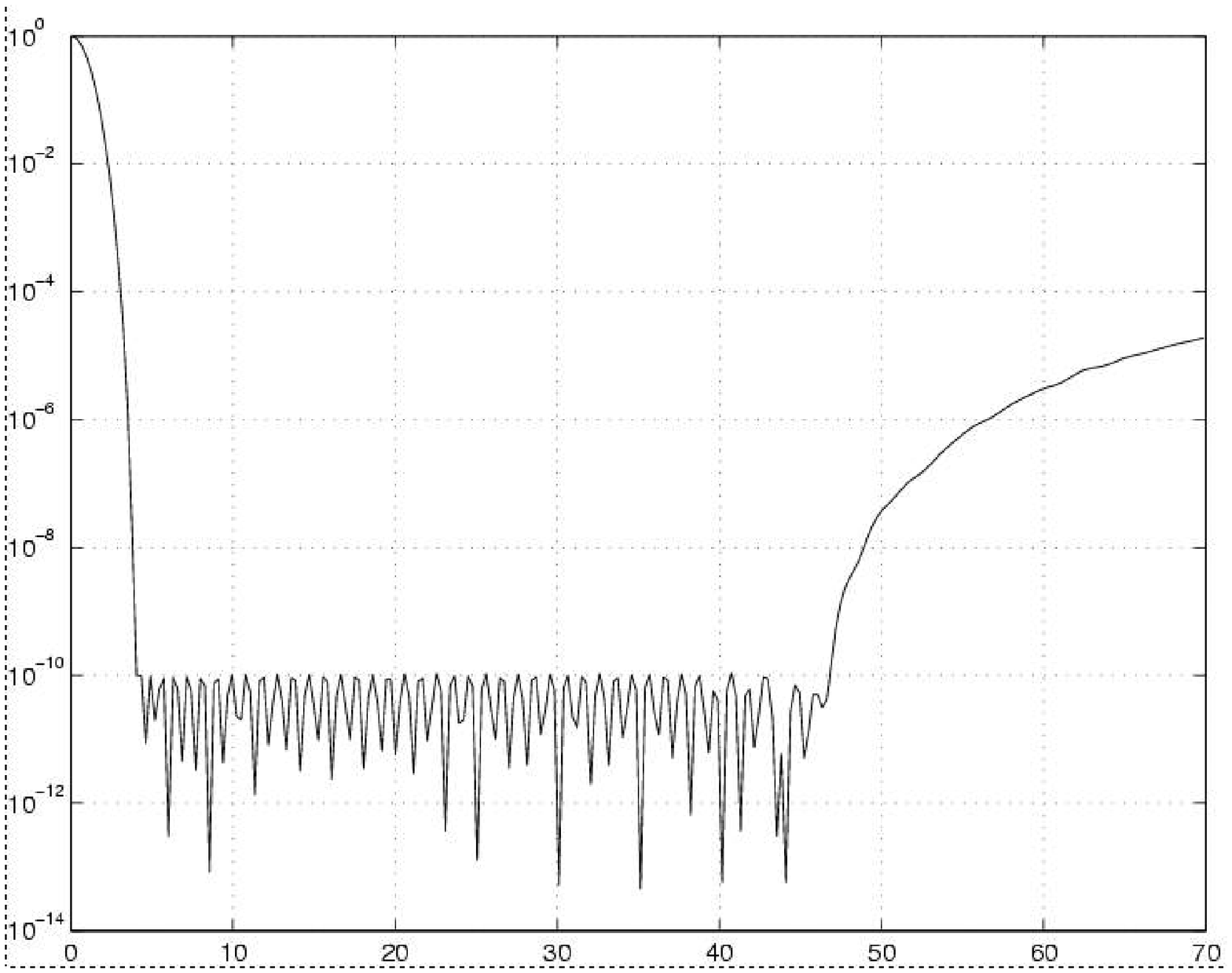}
\hfill ~ 
\end{center}
\caption{Psf's for starshape mask in associated with apodization
shown in Figure \ref{fig:fig7}.
{\em Top Row.} $20$-point star.
{\em Second Row.} $150$-point star.
}
\label{fig:fig2}
\end{figure}

The solution obtained for $\rhonaught = 4$ and $\rhoone = 60$ is shown in 
Figure \ref{fig:fig7}. 
The total-throughput for this apodization is $9.12\%$.  Its Airy-throughput
is $9.09\%$
and its pseudo-area is $17.39\%$.
The psf's for the corresponding starshape
masks with $20$ and $150$ point stars are shown in Figure \ref{fig:fig2}.   
Other apodizations satisfying the smoothness constraints
(such as the generalized prolate spheroidal apodization
discussed in the next subsection), 
being less optimal, must necessarily have even smaller throughputs.

\begin{figure}
\begin{center} 
\includegraphics[width=3.0in]{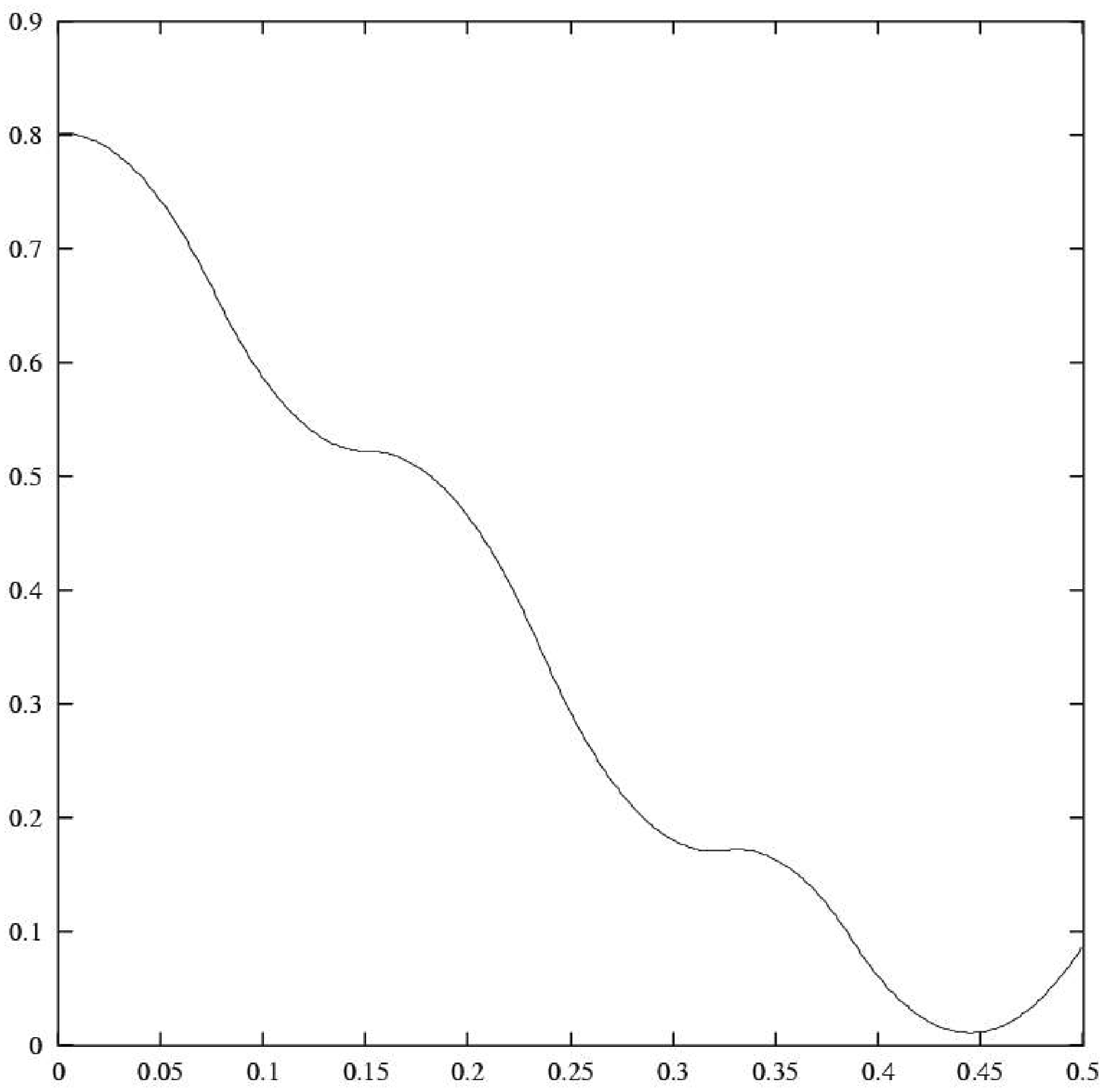}
\includegraphics[width=3.0in]{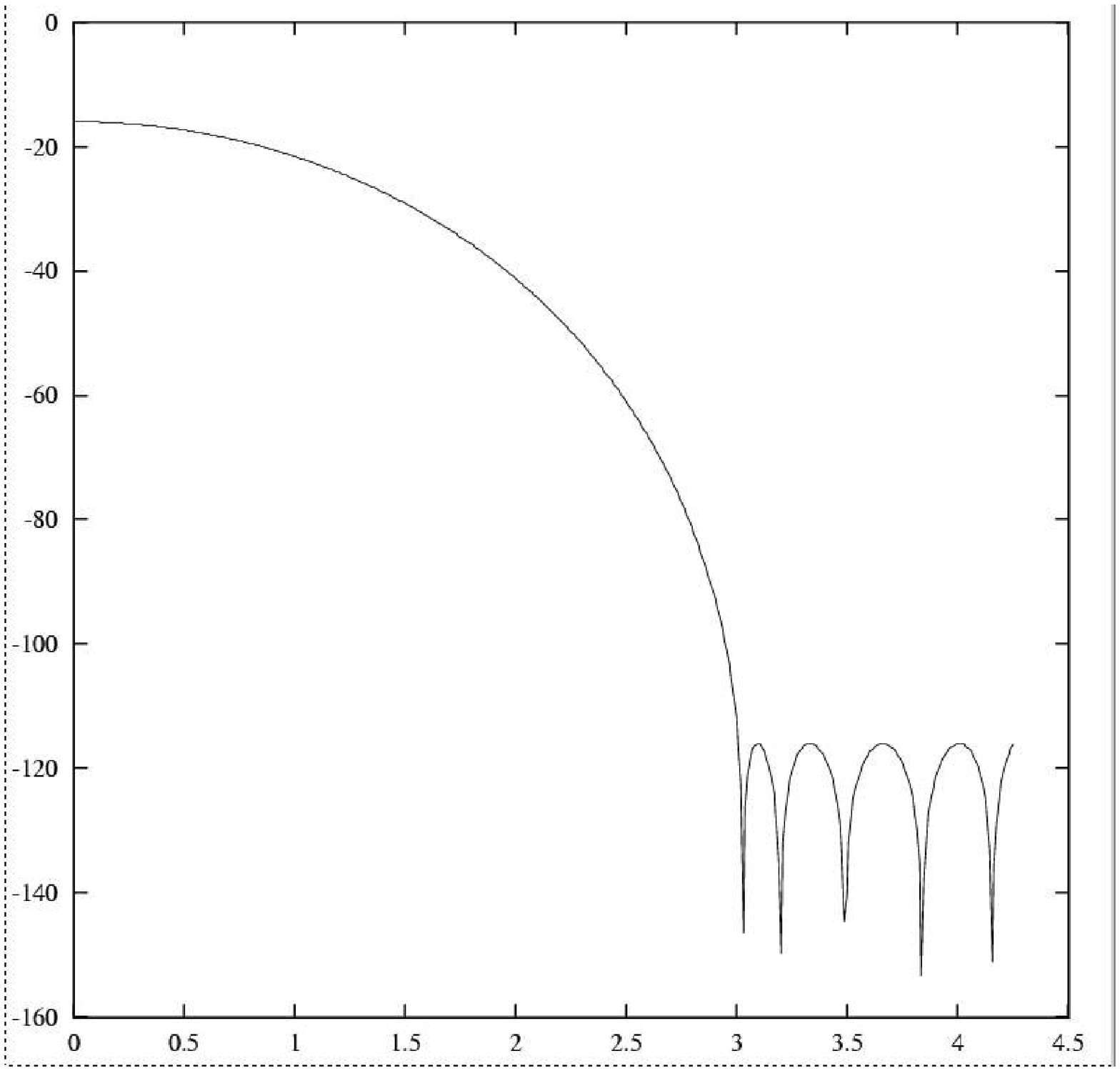}
\end{center}
\caption{The optimal apodization for $\rhonaught = 3$ and $\rhoone = 4.25$
and the associated psf.  Total throughput is $7.7\%$.  Pseudo-area
is $20.0\%$.  Airy-throughput is $7.6\%$.
}
\label{fig:fig4}
\end{figure}

\begin{figure}
\begin{center} 
\hfill
\includegraphics[width=2.0in]{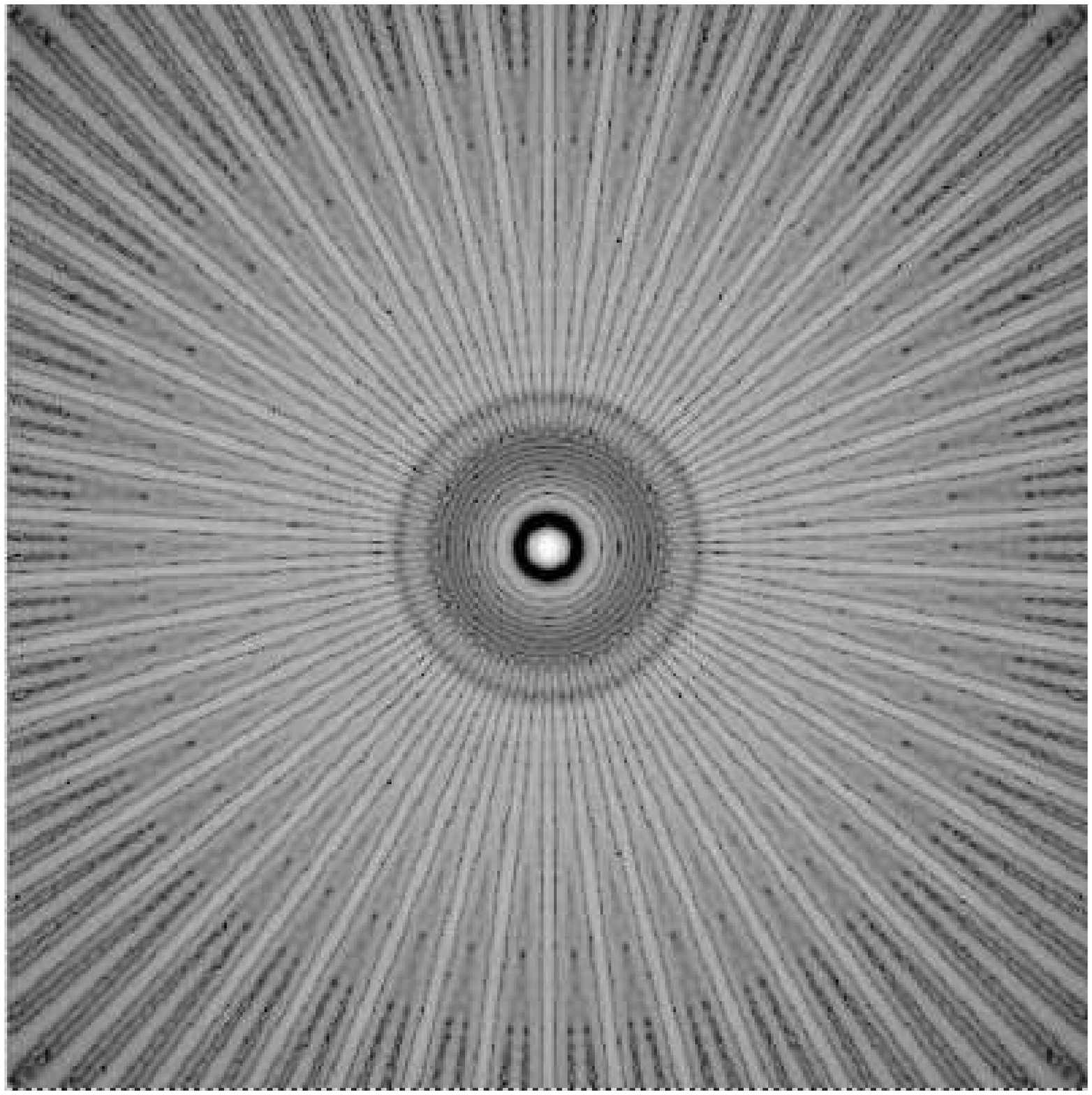}
\hfill
\includegraphics[width=2.5in]{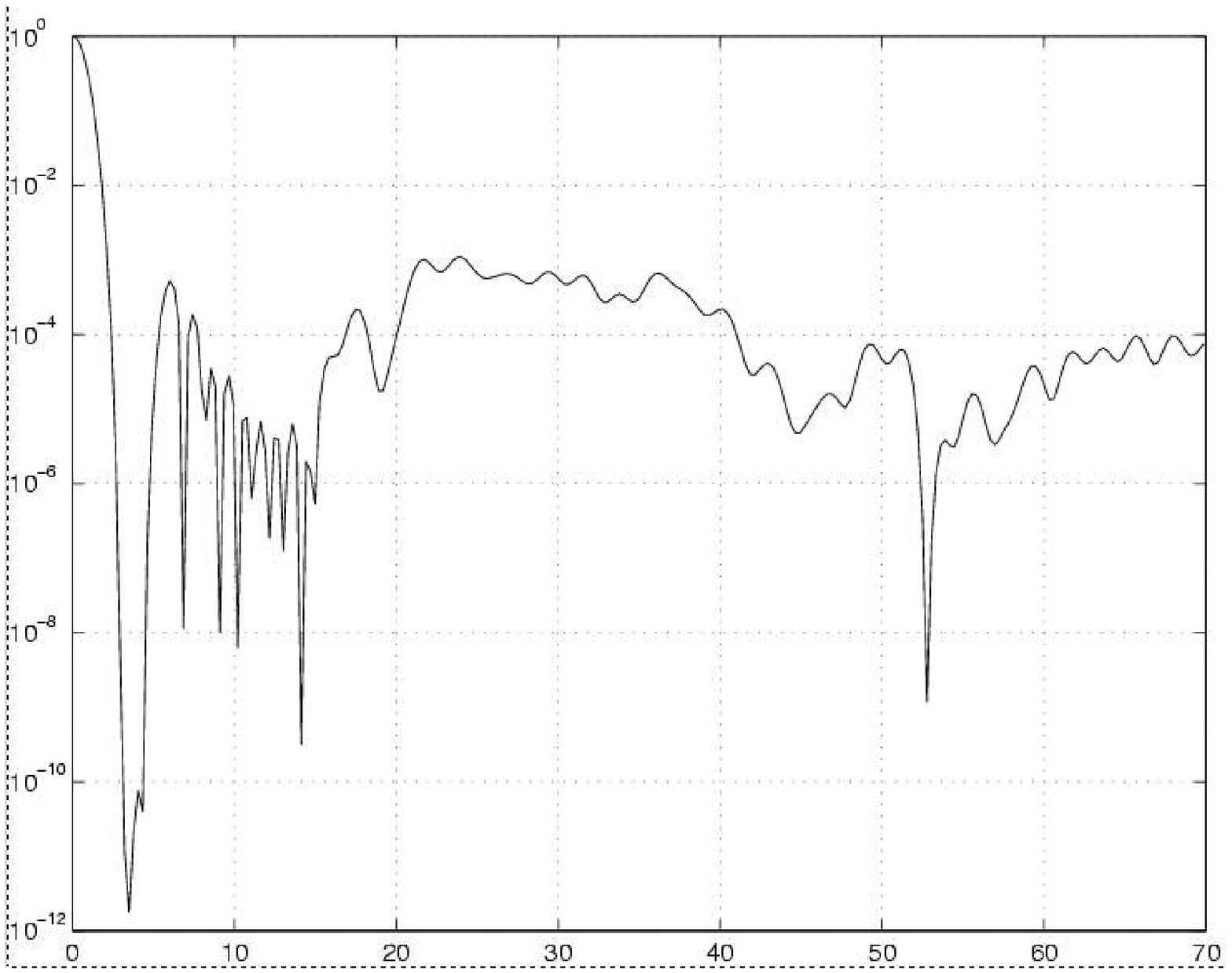}
\hfill ~ 
\end{center}
\caption{Psf's for starshape mask in associated with apodization
shown in Figure \ref{fig:fig4}.
$50$-point star.
}
\label{fig:fig3}
\end{figure}

There is great
interest in reducing the inner working distance to something less than 
$\rhonaught = 4$,
as this would increase the number of likely target stars for TPF. 
Smaller inner working distances can be achieved if
if one is willing to accept a smaller outer working distance too
(\cite{TdF52}).
Unfortunately, a small reduction in inner working distance requires a large
reduction in outer working distance so that very quickly the discovery zone
becomes a very narrow annulus.  For example,
the solution obtained for $\rhonaught = 3$ and $\rhoone = 4.25$ 
is shown in Figure \ref{fig:fig4}.  
The total-throughput for this apodization is $7.7\%$.
Its Airy-throughput is $7.6\%$ and its pseudo-area is $20.0\%$.
The $50$-point starshaped mask associated with this apodization
is shown in Figure \ref{fig:fig3}.

Optimizing pseudo-area or total-throughput are really just simple 
surrogates for the
true objective, which is to minimize integration time.  \citet{ref:Burrows94} 
has shown
that maximizing {\em sharpness} translates directly into minimizing the
integration time required to reach a specified signal to noise ratio.
However, it seems to us that optimizing sharpness is beyond the capabilities
of our optimization tools.  Hence, we optimize the simple surrogates and
reserve sharpness calculations for later comparing various designs.


%
%

\subsection{The Generalized Prolate Spheroidal Apodization}

\begin{figure}
\begin{center} 
\includegraphics[width=3.0in]{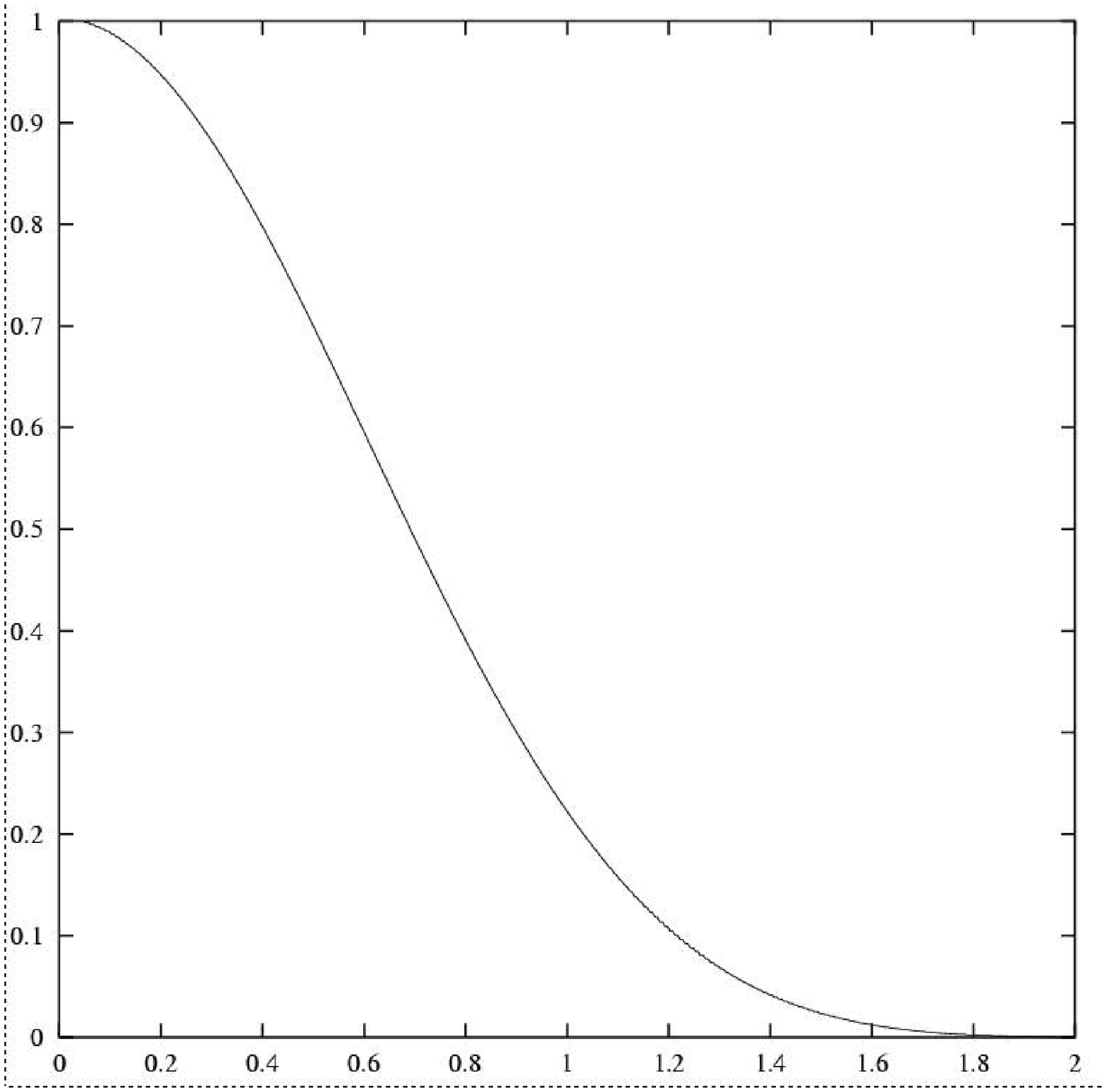}
\includegraphics[width=3.0in]{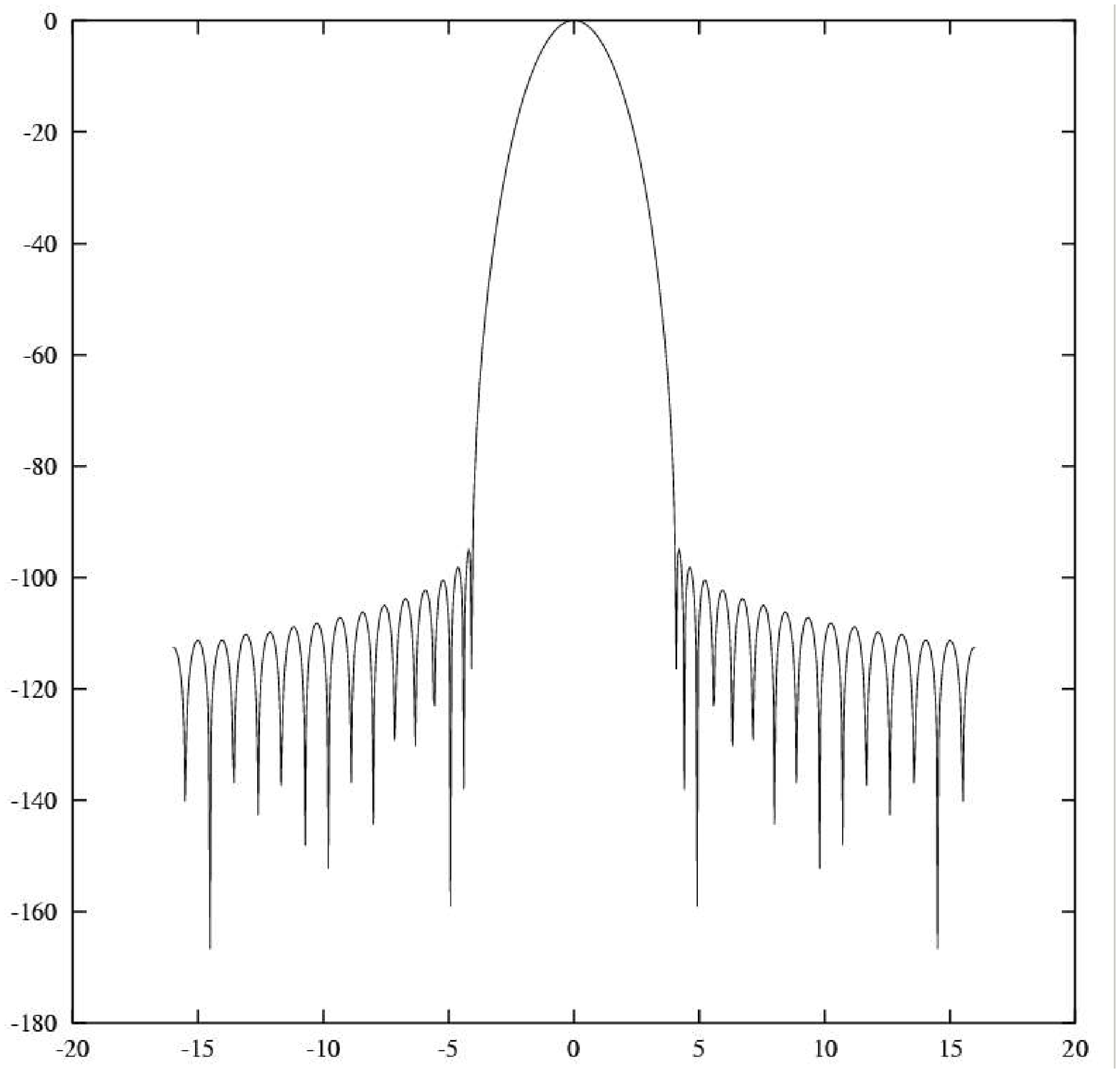}
\end{center}
\caption{{\em Left.} The generalized prolate spheroidal apodization computed
using $\rhonaught = 4$.  Total throughput and Airy-throughput agree and are
equal to $8.52\%$.  The pseudo-area is $16.28\%$.
{\em Right.} The associated psf.
}
\label{fig:fig5}
\end{figure}

Generalized prolate spheroidal wave functions were first introduced
by \citet{ref:Slepian} as a way to apodize a circularly symmetric 
aperture so as to concentrate as much light 
as possible into a central Airy disk.  
In the context of TPF, the generalized
prolate spheroidal wave function has been
popularly viewed as providing an optimal apodization
(see, e.g., \cite{KVSL02,ASF01,GN02}).
However, in truth, 
it is an optimal solution to the following slightly different 
optimization problem\footnote{Slepian actually formulated it 
as an equivalent unconstrained optimization problem: 
$\max \int_0^{\rhonaught} E(\rho)^2 d\rho / \int_0^{\infty} E(\rho)^2
d\rho$.}:
\begin{equation}
    \begin{array}{ll}
        \mbox{minimize } & 
	\displaystyle \int_{\rhonaught}^{\infty} E(\rho)^2 d\rho \\[0.2in]
	\mbox{subject to } &
	    \setlength{\arraycolsep}{0.1em}
	    \begin{array}[t]{rcll}
	    \displaystyle
		& A(0) & = 1 .
	    \end{array}
    \end{array}
\end{equation}
The generalized prolate spheroidal apodization computed using $\rhonaught = 4$
gives $8.3\%$ total-throughput.  The Airy-throughput is essentially the same.
The pseudo-area is $15.9\%$.
The apodization and corresponding 
psf are shown in Figure \ref{fig:fig5}.
At the expense of 
Airy-throughput ($8.3\%$ vs. $9.1\%$) and
slight violation of the contrast in the first few diffraction rings,
it has better than needed contrast throughout most of a larger than 
needed dark zone.

The throughputs for the apodizations discussed in this section are
summarized in Table \ref{tbl1}.

\section{Final Remarks}

Some TPF concepts involve an elliptical pupil geometry since this might
provide a means to achieve improved angular resolution in realizable
rocket fairings.  The designs presented in this paper 
are given in unitless variables.  When re-unitizing, a different
scale can be used for the $x$ and $y$ directions.  In this way, these
designs can be applied directly to elliptical pupils.  Of course, the 
high-contrast region of the psf will also be elliptical with the short
axis of the psf corresponding to the long axis of the pupil.

There are two parts to TPF: discovery and characterization.  Discovery refers
to the simple act of looking for exosolar planets.  Characterization refers to
the process of learning as much as possible about specific planets after they
have been discovered.  The masks presented here are intended primarily for
discovery since a single exposure with these masks
can discover a planet in any orientation relative to its star.  
However, once a planet is found and its orientation is known,
some of the asymmetric masks presented in previous papers should be used for
photometry and spectroscopy as they have significantly
higher single-exposure throughput.


An alternative to pupil masks is to use a traditional coronagraph, which
consists of an image plane mask followed by a Lyot stop in a reimaged pupil
plane.  Recently, \citet{ref:kuchner} have developed band-limited image-plane
masks that achieve the desired contrast to within $3 \ld$.  A potential
drawback of this 
approach is its sensitivity to pointing accuracy.  Nonetheless, it
appears very promising.  In the future, we plan to consider combining pupil
masks with image masks and Lyot stops to make a hybrid design that hopefully
will provide a design achieving the desired contrast with
an even smaller inner working distance and perhaps higher throughput.

In this paper we have only considered scalar electric fields.  We leave
the important and more complex issue of how to treat vector 
electric fields, i.e.  polarized light, to future work.

{\bf Acknowledgements.}
We would like to express our gratitude to our colleagues 
on the Ball Aerospace and Technology TPF team.  
We benefited greatly from the many enjoyable and stimulating
discussions.  
This work was partially performed for the Jet Propulsion Laboratory, California
Institute of Technology, 
sponsored by the National Aeronautics and Space Administration as part of
the TPF architecture studies and also under contract number 1240729.
The first author received support from the NSF (CCR-0098040) and
the ONR (N00014-98-1-0036).

\bibliography{../lib/refs}   
\bibliographystyle{plainnat}   

\clearpage

\clearpage

\begin{table}
\centering
\begin{tabular}{||l|c|c|c||c|c|c||} \tableline
\emph{} & \emph{Smooth?} & \emph{$\rhonaught$} & \emph{$\rhoone$} & 
    \emph{${\mathcal T}_{\rm{total}}$} &
    \emph{${\mathcal T}_{\rm{Airy}}$} &
    \emph{$E(0)$}
    \\
\tableline
Figure \ref{fig:fig10} & No  & 4 & 60   & 17.9\pz & 9.37   & 17.9\pz \\
Figure \ref{fig:fig7}  & Yes & 4 & 60   &  9.12   & 9.09   & 17.39 \\
Figure \ref{fig:fig4}  & Yes & 3 & 4.25 &  7.7\pz & 7.6\pz & 20.0\pz \\
Figure \ref{fig:fig5}  & Yes & 4 & 60   &  8.34   & 8.34   & 15.94   \\
\tableline
\end{tabular}
\caption{Summary of the throughput results for the specific apodizations
considered.}
\label{tbl1}
\end{table}

\end{document}